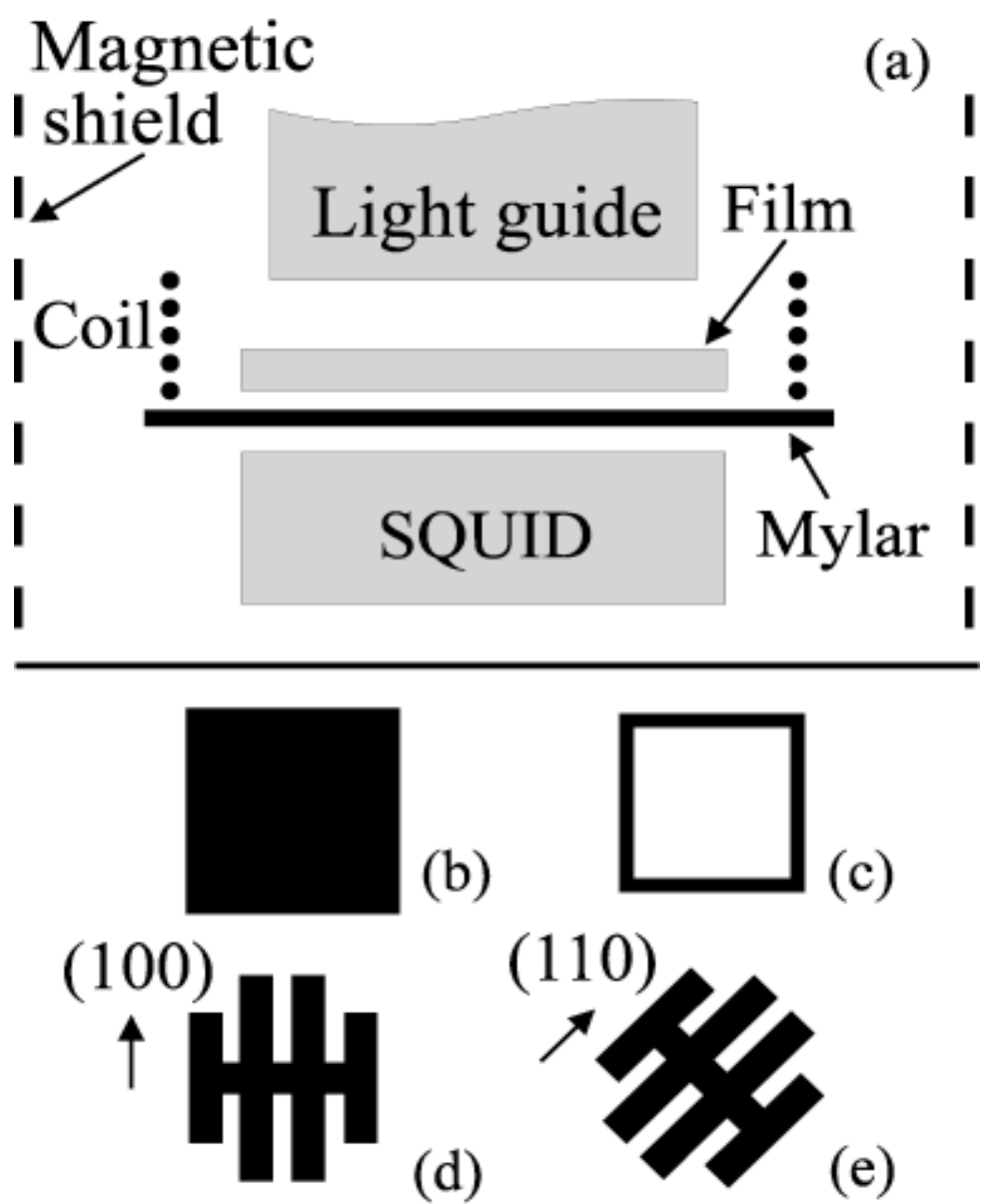

Fig. 1

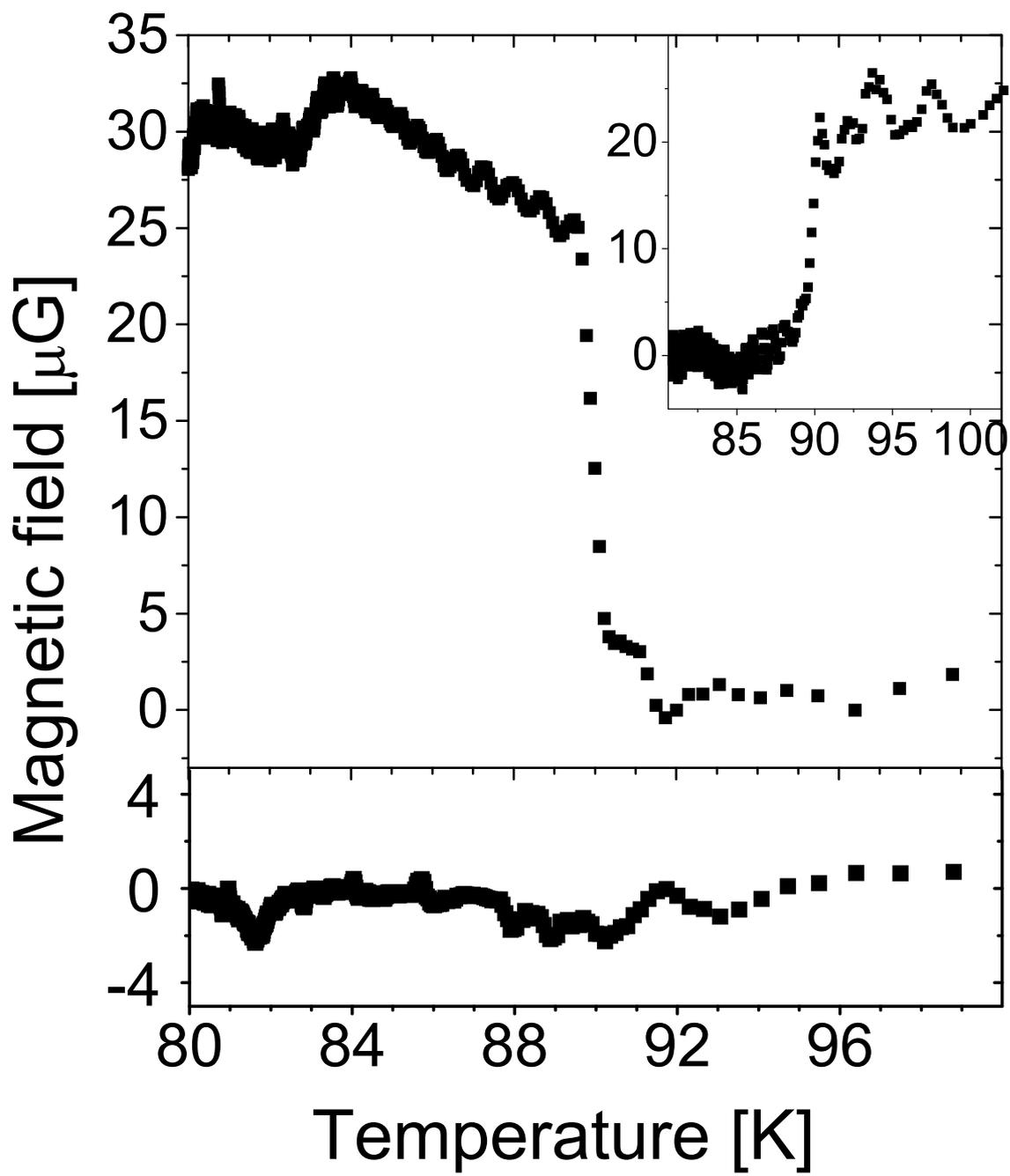

Fig. 2

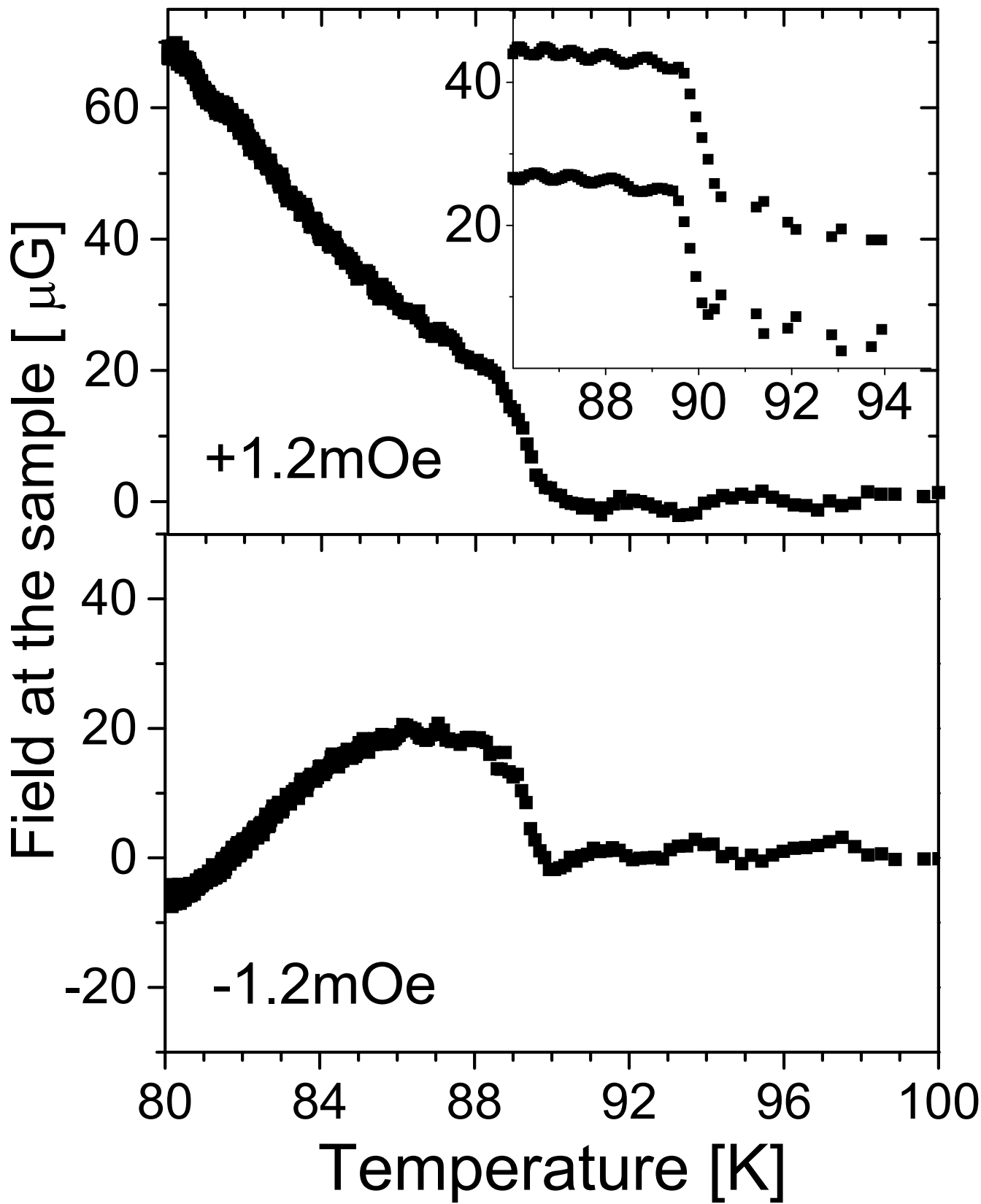

Fig. 3

# Appearance of Spontaneous Macroscopic Magnetization at the Superconducting Transition of Yba$_2$Cu$_3$O$_{7-\delta}$

by

R. Carmi, E. Polturak[*] G. Koren, and A. Auerbach

Physics Department
Technion - Israel Institute of Technology, Haifa 32000, ISRAEL

One of the most fascinating aspects of high temperature superconductivity is the unconventional symmetry of the order parameter. Several experiments[1-3] established that the order parameter has a $d_{x2-y2}$ symmetry under rotation of the lattice. An intriguing, and much debated possibility is that an additional, imaginary component may be present in certain cases, having an isotropic s-wave[4-6] or $d_{xy}$ symmetry[7-10]. A complex order parameter of the type $d_{x2-y2} + id_{xy}$ breaks both parity (P) and time reversal symmetry (T). A clear signature of both P and T symmetries being broken in the superconducting state would be the spontaneous appearance of a macroscopic magnetization, as in a ferromagnet. Broken T symmetry has been reported[5,15], while search for effects related to combined P and T breaking gave null results[11-14]. Here, we report the observation of a weak (~ 10$^{-5}$ G) spontaneous magnetic field appearing at the transition temperature of Yba$_2$Cu$_3$O$_{7-\delta}$ into the superconducting state. The magnetic signal originates near the edges of the epitaxial thin film samples. We offer two alternative interpretations: (i) The condensate carries an intrinsic angular momentum, and a magnetic moment between 10$^{-2}$ $\mu_B$ and 10$^{-4}$ $\mu_B$ per plaquette. (ii) π junctions which may exist near the edges produce circulating supercurrents and magnetic flux. Our data imposes constraints on each of these scenarios.

Previous experimental searches of combined P and T violation set a limit of a few percent on any symmetry breaking component of the order parameter[11-14]. If a spontaneous magnetic field below this limit were to exist, it may be easier to detect it by looking at the magnetic flux produced by the whole sample, instead of a small region. This is conditional upon such field having the same orientation everywhere in the superconductor. To check this possibility, in our experiment we placed high quality epitaxial, c-axis oriented Yba$_2$Cu$_3$O$_{7-\delta}$ films atop an input coil of a HTSC dc-SQUID magnetometer (see Fig. 1a). The magnetometer (M2700L made by Conductus, Inc.,) has a large, 8mm×8mm directly coupled single input loop. The magnetometer is operated in a flux locked loop, with either ac or dc bias. Films of YBa$_2$Cu$_3$O$_{7-\delta}$ were prepared either by Laser Ablation Deposition or DC Sputtering on 1cm×1cm size substrates, including (100) SrTiO$_3$, (100)MgO and (001)NdGaO$_3$. The range of thickness was between 30 nm and 300 nm, with T$_c$ typically around 90 K. The films were measured as deposited, or after patterning into different structures described below. We have also tested pressed YBCO powder samples. In total, we measured 15 films, of which 14 showed spontaneous flux. Hence, the effect described here does not depend on the growth method or the substrate. In our measurement setup shown in Fig. 1a, the distance betwen the sample and SQUID is 1mm. Despite the proximity, the film and the SQUID are located in two different chambers separated by a mylar membrane, which allows us to vary the temperature of the films while keeping that of the SQUID constant at the base temperature of 77 K. Magnetic shields reduce the residual field down to 10$^{-4}$ G. Additional coils are used to further reduce this field, or to check for any field dependence. In order to avoid stray fields generated by currents used in resistive heating, the films are heated by a guided light beam,. To eliminate any thermoelectric currents, the sample holder and all nearby components were made of non magnetic plastic. Cooling of the samples is done using He exchange gas. We verified, by changing the cooling rate by two orders of magnitude (K/sec to K/min) that the spontaneous signal was independent of thermal gradients. Temperature of the films is measured in-situ using the resistance of a carbon film painted onto the substrate. We verified that the small AC current used to measure the thermometer does not affect the results.

Fig. 2 shows a typical signal recorded by the magnetometer during the cooling of a film through $T_c$. A small spontaneous magnetic field appears abruptly at $T_c$, increasing in magnitude over an interval of about 1K, and saturating below it. The ~1K interval is typical of the spread of $T_c$ across the wafer. Hence, the true interval is less than 1K. The bottom part of Fig. 2, shows the reference measurement carried out using a blank substrate. In order to ascertain that the effect is not caused by partial screening of the inductance of the SQUID as the sample cools through $T_c$, we inverted the film relative to the SQUID, and found that the polarity of the spontaneous signal was reversed (see inset in Fig. 2). A reversal of the signal rules out screening as a source of the effect. As an additional check, we measured the screening properties of the various film patterns (Fig. 1b-1e) using a low temperature inductance bridge. We found no correlation between the spontaneous signal and the screening properties. For example, the pressed powder samples had negligible screening relative to that of a film, while the spontaneous signal was 10-100 times larger. In addition, we ascertained that the bias polarity and the magnitude of the current in the feedback loop of the SQUID did not effect the signal (see inset in Fig. 3).

It is equally important to rule out external magnetic fields as a source of the effect. Any such field would be partially expelled from the film below $T_c$, inducing a signal proportional to the external field both in magnitude and sign. We repeated the measurements in the presence of magnetic fields up to 100 times larger than the residual field and along different directions. In a finite field, a temperature dependent background appears below $T_c$, in addition to the spontaneous signal. An example of such measurement is shown in Fig. 3. Evidently, reversal of the external field between Fig. 3a and 3b changes the sign of the background, but not of the spontaneous signal such as shown in Fig. 2. The latter remains superimposed on this background, independent of both the magnitude and direction of the external field. This rules out the possibility that the effect originates from the presence of external fields or internal magnetic impurities. Comparing the data of Fig. 2 with that of Fig. 3, the weak temperature dependence seen below $T_c$ in Fig. 2 is consistent with the presence of residual field of ~$10^{-4}$ G. Consequently, the variation of the *spontaneous* field between $T_c$ and 77 K is less than 10% of the jump at $T_c$.

Geometry: in Fig. 4 we plot the size of the spontaneous field jump vs. film thickness *d*. The comparison is made for unpatterned films(Fig. 1b), with identical areal dimensions. It is seen that within the scatter, the signal does not depend on *d*. In the superconducting state, any magnetic moment appearing in the bulk of the film must be shielded by the Meissner effect. Net magnetic signal should therefore originate only from the edge of the film. We repeated the measurement after removing most of the film by litography leaving only a loop (Fig 1c). The magnitude of the signal coming from the loop did not change appreciably relative to the original film, which indicates that the magnetic signal originates near the edges. We also tested films patterned so that their edges were oriented along different crystalline directions, (100) or (110)(see Fig. 1d and 1e). We found that the signal did not depend on the orientation of the pattern, but in general, the magnitude of the signal increased with the the length of the perimeter. This conclusion is further supported by data taken with pressed powder samples having much larger surface area than the films. In this case, the spontaneous signal was 10-100 times larger than shown in Fig. 2.

Discussion: Our findings are summarized as follows: First, after the initial rise below $T_c$, the field picked up by the SQUID is independent of both temperature and film thickness. Second, the origin of the field is near the edges of the film. We examine two different scenarios; (a) the effect is caused by an intrinsic angular momentum of the superconducting order parameter. (b) The film contains defects near its edges which behave as internal π Josephson junctions.

Regarding scenario (a), we denote the intrinsic magnetization density by $m_z$. We assume that the flux emanates from a region of width $W$ near the edge of perimeter length $L$. The flux measured by the SQUID is given by:

$$\Phi = \alpha\, 4\pi\, m_z\, W L \qquad (1)$$

here, $\alpha$ is flux coupling coefficient between the sample and the SQUID. We determined $\alpha$ at about 0.5. Several authors[4,9] predict that magnetization related to P and T breaking appears only within a coherence length $\xi$ near defects or at surfaces, i.e. $W = \xi$. There is no dependence on $d$. In order for these theories to explain also the temperature dependence of $\Phi$, they should predict $m_z(T) \propto 1/\xi(T)$ which is also proportional to the gap $\Delta$. For the data of Fig. 2, the flux generated by the sample is $7.7 \times 10^{-6}$ G cm$^2$, or 37 $\phi_0$. Extrapolating to T=0, and using $\xi_0 = 15$Å, we find the edge field (inside a width $\xi_0$) would be 23 G. Fields of this magnitude were suggested in Ref. 9.

Another possibility is that the intrinsic magnetization is a property of the bulk[7,8]. However, a non-zero magnetic signal would come only from a region of width $W = \lambda_{eff}$ which is not shielded by the Meissner effect[7]. In the case of a thin film, $\lambda_{eff} = \lambda^2/d$, where $\lambda$ is the London penetration depth[16]. This relation holds for $d < \lambda$ which is always true near $T_c$. In this case, $m_z(T)$ should be proportional to $d/\lambda^2$ to cancel the temperature and thickness dependence of $\Phi$. Thus, to be in line with our data, the predicted $m_z(T)$ should be proportional to the areal superfluid density. Extrapolating to $T = 0$, and using $\lambda_0 = 1400$Å, and $d = 2500$Å, we find the edge field (inside a width $\lambda_0^2/d$) would be 0.37G for the sample of Fig. 2. If one looks at the edge area using a SQUID microscope with a 10μm × 10 μm pickup loop[17], the average field sensed would be about $10^{-3}$G for both cases ($W = \xi_0$ or $W = \lambda_0^2/d$). Edge fields of this magnitude were indeed observed[17].

From the results, we set limits on the intrinsic magnetic moment per plaquette. If $W = \xi_0$, we calculate from Eq.(1) a magnetic moment of $1.8 \times 10^{-2}$ $\mu_B$ per CuO$_2$ plaquette. In the case of $W = \lambda_0^2/d$, we get $3.7 \times 10^{-4}$ $\mu_B$ per CuO$_2$ plaquette. If one further assumes that $m_z$ reflects the intrinsic orbital moment density of the Cooper pairs, we can convert it to an intrinsic angular momentum per Cooper pair $<L_z>$. Assuming a pair density of 0.08 per CuO$_2$ plaquette (half of the hole density), and an orbital magnetic moment of 4 $\mu_B <L_z>/$ per pair, we obtain $<L_z>/$ 1.2×10$^{-3}$ if $W = \lambda_0^2/d$ and $<L_z>/ = 6 \times 10^{-2}$ if $W = \xi_0$. This implies $\Delta_{xy}/\Delta_{x2-y2} \approx 5 \times 10^{-4}$ for $W = \lambda_0^2/d$ and $\Delta_{xy}/\Delta_{x2-y2} \approx 2.5 \times 10^{-2}$ if $W = \xi_0$. Thus, the $d_{xy}$ component is very small, which may explain why it went so far undetected.

Turning to the second scenario, it is well known that spontaneous flux of $\phi_0/2$ appears in a superconducting loop containing a π junction[1,2,18,19]. The flux coming from the sample in Fig. 2 could be generated by 74 junctions with circulating currents all in the same sense. This would explain the jump at $T_c$, temperature and thickness independence below $T_c$. The fact that the signal originates near the edges implies that such junctions are perhaps associated with defects near the boundary. There are however several points which are unclear regarding this interpretation. First, large angle grain boundaries which are usually used to construct π junctions[2] are absent in epitaxial films. Thus, one has to look for defects of another kind. Second, removing the original boundary of the film by patterning did not change the signal size appreciably. Third, the total flux measured seems to be the same in different films (Fig. 4). This requires that the number of spontaneously created vortices and their sign is not random.

Finally, the inset in Fig. 2 and the main panel of Fig. 3 show that the direction of the spontaneous field is robust with respect to small external fields and temperature cycling. However, this direction, altough seemingly fixed for each sample, was random between different samples. This suggests that the direction of this field is determined at temperatures above room temperature, certainly higher than $T_c$ .

**Acknowledgments:** We acknowledge useful discussions with E. Zeldov. This work was supported by the Israel Science Foundation, The Heinrich Hertz Minerva Center for HTSC, and by the VPR Technion Fund for Promotion of Research.
Correspondence should be addressed to E. P. (e-mail: emilp@physics.technion.ac.il )

# Figure captions

Figure.1 Schematic illustration of the experimental geometry and of the sample patterns. (a) cross section of the experimental setup. Light is introduced briefly through the light guide to heat the sample above $T_c$, and is turned off during the measurements. (b)-(e) film patterns used in this work.

Figure 2. Spontaneous magnetic field generated by a thin YBCO film vs.temperature. The sample is patterned as a disc, 0.7 cm in diameter. The inset shows the signal measured after the film was inverted with respect to the SQUID. The magnitude of the signal both in the figure and in the inset is corrected for the different SQUID-sample distance, 1 mm in the usual configuration and 2 mm with the film inverted. The bottom part shows a reference measurement, done with a blank substrate.

Figure 3. Total signal generated by a YBCO film cooled in an external magnetic field. The external field, 1.2mG, is an order of magnitude larger than the residual field. Altough the external field is reversed between the top and bottom parts of the figure, the spontaneous part of the signal appearing at $T_c$ remains unchanged. The inset shows the spontaneous signal at zero field obtained with reversed polarities of the SQUID bias. For clarity, the data were offset by a constant value.

Figure 4. Magnitude of the spontaneous field jump vs. film thickness. The various symbols in the figure are used only for sample identification. Error bars reflect the noise level in the data, while the dashed line is the average value of the signal.

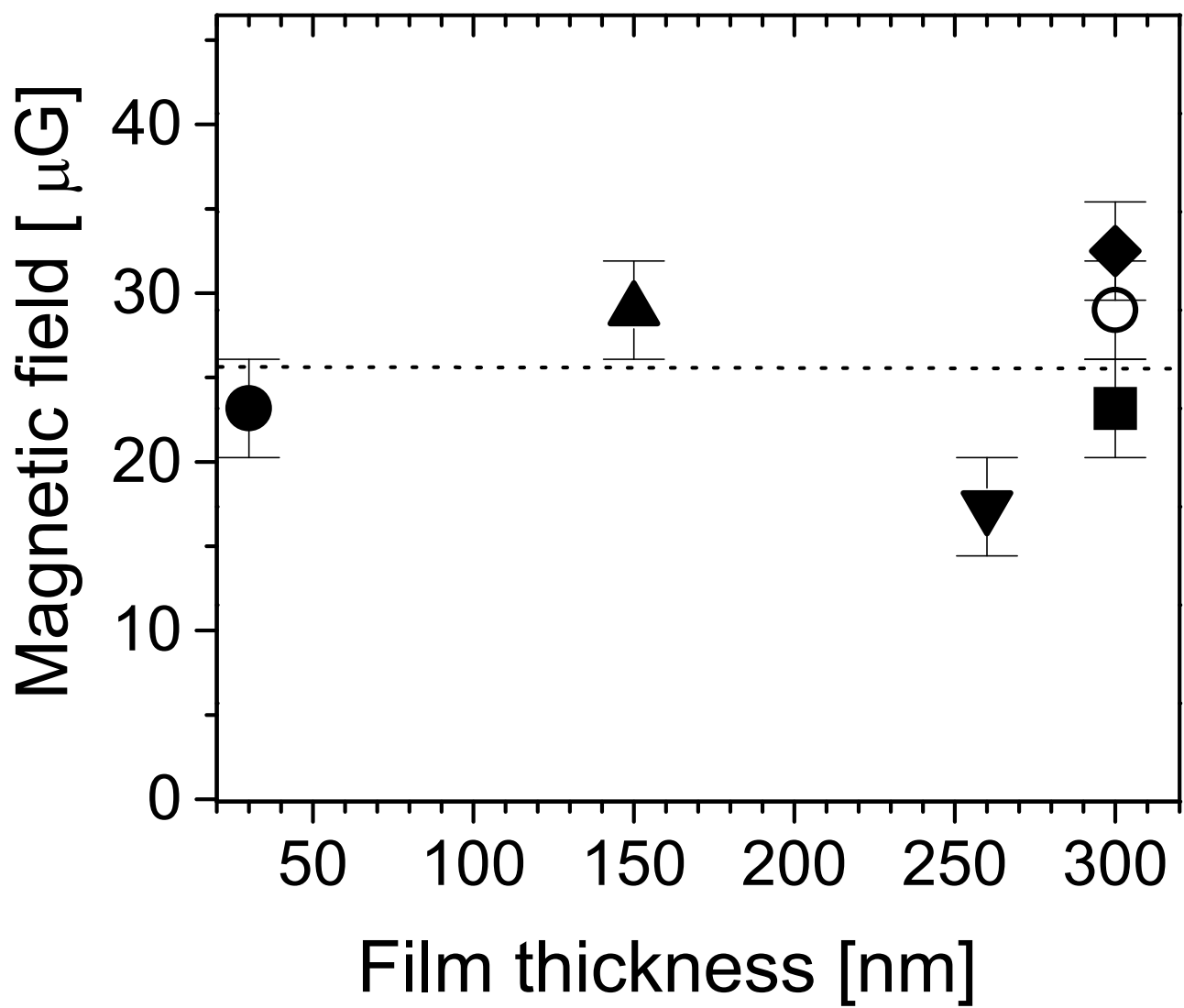

Fig. 4